\documentstyle[11pt,newpasp,twoside,psfig]{article}
\index{pulsars!B1509-58}
\index{pulsars!braking indices}
\index{pulsars!spin-down}
\index{timing}
\index{timing!phase-coherent}

\newcommand{\psr}{PSR~B1509$-$58}
\newcommand{\nudotdotdot}  {\stackrel{\bf{\ldots}}{\nu}}

\def\edcomment#1{\iffalse\marginpar{\raggedright\sl#1\/}\else\relax\fi}
\reversemarginpar

\begin{document}

\title{Long-term Radio Timing Observations of \psr}
\author{Margaret A. Livingstone, Victoria M. Kaspi}
\affil{McGill University, 3600 University Ave, Montreal, Quebec, Canada. H3A
2T8.}
\author{Richard N. Manchester}
\affil{Australia Telescope National Facility, CSIRO, P.O. Box 76, Epping NSW
1710, Australia}

\begin{abstract}
We present an updated phase-coherent timing solution for the young, energetic
pulsar \psr\ for twenty years of data. Using a partially phase-coherent
timing analysis, we show that the second frequency derivative is changing in
time implying a third frequency derivative of 
$\nudotdotdot= (-9 \pm 1) \times 10^{-32} $s$^{-4}$. This value is consistent
with the simple power law model of pulsar rotation. 
\end{abstract}

\section{Introduction}
Radio pulsars are powered by rotational kinetic energy and experience losses
due to the emission of electromagnetic energy while spinning down. 
The spin-down of radio pulsars is described by (Manchester \& Taylor, 1977)
\begin{equation}
\dot{\nu} \propto -{\nu}^n , 
\end{equation}
where $\nu$ is the spin frequency, $\dot{\nu}$ is the frequency derivative and 
$n$, the braking index, equals 3 for a spinning magnetic dipole in a vacuum.
Taking a derivative of (1), we find that $n = \nu \ddot{\nu}/{\dot{\nu}^2}$.
Taking further derivatives of (1), 
and defining the second braking index as $m_0=n(2n-1)$\cite{Kaspi}, 
we find an expression $m = {{\nu^2} \nudotdotdot}/{\dot{\nu}^3}$. 
Measuring a value of $m$ that
agrees with the predicted value $m_0$ is an excellent check on the validity
of the power law spin-down model of radio pulsars given in equation (1). 
Kaspi et al. (1994) reported a value of $n=2.837 \pm 0.001$ 
and $m=14.5 \pm 3.6$ for \psr. 

\section{Results and Discussion}
Performing a phase-coherent timing analysis on twenty years of radio timing
data for \psr\ obtained from the Molongolo Observatory Synthesis Telescope and 
the Parkes 64m telescope, we found the second frequency derivative to be
$\ddot{\nu} = 1.962(1) \times 10^{-21}$~s$^{-3}$, implying a braking index
of $n = 2.814(1)$. Significant
low-frequency timing noise in the data prohibited a phase-coherent analysis
to determine an accurate value of the third frequency
derivative. This prompted a partially phase-coherent timing analysis by breaking up
the data into subsets of $\sim 2$ yr (such that phase residuals were `white'
in each interval following a fit of $\nu$, $\dot\nu$ and $\ddot\nu$) 
and plotting the values of
$\ddot{\nu}$ against time, the results of which are shown in Figure 1. This method is
less sensitive to timing noise which dominates on long time scales. We found
a value $\nudotdotdot = (-9 \pm 1) \times 10^{-32} $s$^{-4}$ by a weighted
least-squares fit to the data. This value for
$\nudotdotdot$ is in agreement with that predicted from the spin-down law
of $\nudotdotdot = -9.37 \times 10^{-32} $s$^{-4}$. This measured value for
$\nudotdotdot$ implies a second braking index of $m= 12.7 \pm 1.4$, also in
agreement with the predicted value of $m=13.3$, though less than the value
$m=15$ expected for an ideal spinning dipole. 
Our measurement of the $\nudotdotdot$ implies that the spin-down law correctly
predicts the behaviour of \psr, reinforcing the use of the power law given in 
equation (1) to describe the rotation of pulsars. 

\begin{figure}
\begin{center}
\centerline{\psfig{file=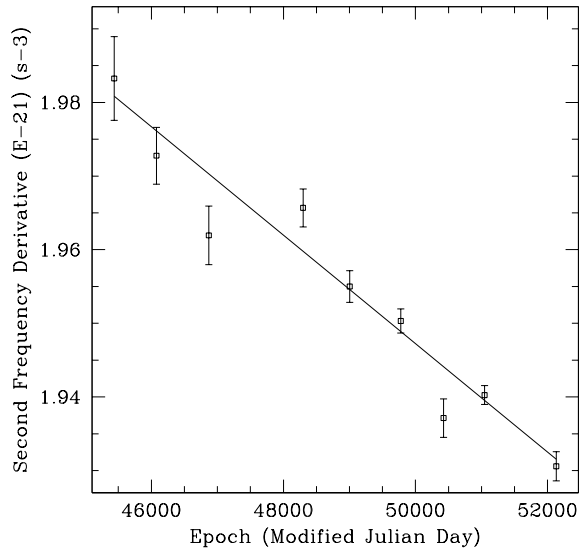,width=8cm}}
\end{center}
\caption{Phase-coherent subsets plotted versus epoch. The slope of the 
line is the third derivative and has a value of $(-9 \pm 1) \times 10 ^{-32} $s$^{-4}$. }
\end{figure}

\end{document}